\documentclass[gmd, manuscript]{copernicus}
\nolinenumbers

\usepackage[section]{placeins}
\usepackage{tabularx}

\usepackage{graphicx}
\usepackage{caption}
\usepackage{float}
\graphicspath{{figs/}}

\begin{document}
\title{Machine Learning Parameterization of the Multi-scale Kain-Fritsch (MSKF) Convection Scheme and stable simulation coupled in WRF using WRF-ML v1.0}


\Author[1]{Xiaohui}{Zhong}
\Author[2]{Xing}{Yu}
\Author[1]{Hao}{Li}

\affil[1]{Fudan University, Shanghai, 200433, China}
\affil[2]{Shenzhen Institute of Artificial Intellegence and Robotics for Society, Guangdong, 518000, China}




\correspondence{Xing Yu (yuxing@cuhk.edu.cn)}

\runningtitle{Machine Learning Parameterization of the Multi-scale Kain-Fritsch (MSKF) Convection Scheme}

\runningauthor{X. Zhong et al.}

\received{}
\pubdiscuss{} 
\revised{}
\accepted{}
\published{}


\firstpage{1}

\maketitle

\begin{abstract}
Warm-sector heavy rainfall often occurs along the coast of South China, and it is usually localized and long-lasting, making it challenging to predict. High-resolution numerical weather prediction (NWP) models are increasingly used to better resolve topographic features and forecast such high-impact weather events. However, when the grid spacing becomes comparable to the length scales of convection, known as the gray zone, the turbulent eddies in the atmospheric boundary layer are only partially resolved and parameterized to some extent. Whether using a convection parameterization (CP) scheme in the gray zone remains controversial. Scale-aware CP schemes are developed to enhance the representation of convective transport within the gray zone. The multi-scale Kain-Fritsch (MSKF) scheme includes modifications that allow for its effective implementation at a grid resolution as high as 2 km. In recent years, there has been an increasing application of machine learning (ML) models to various domains of atmospheric sciences, including the replacement of physical parameterizations with ML models. This work proposes a multi-output bidirectional long short-term memory (Bi-LSTM) model as a replace the scale-aware MSKF CP scheme. The Weather Research and Forecast (WRF) model is used to generate training and testing data over South China at a horizontal resolution of 5 km. Furthermore, the WRF model is coupled with the ML based CP scheme and compared with WRF simulations with original MSKF scheme. The results demonstrate that the Bi-LSTM model can achieve high accuracy, indicating the potential use of ML models to substitute the MSKF scheme in the gray zone.
\end{abstract}

\introduction  
Warm-sector heavy rainfall often occurs during the pre-flood seasons in South China due to the influence of the East Asian summer monsoon \citep{ding2004l}. These rainfall events are localized and are characterized by high precipitation intensity but limited spatial coverage. However, despite being small size, unexpected and extreme warm-sector rainfall can cause significant damage, resulting in flooding homes and vehicles, destroying crop fields, and endangering lives, with costs amounting to millions or even billions of dollars \citep{Tao1981,zhao2007onset,zhongetal2015}. Accurately predicting warm-sector heavy rainfall using Numerical Weather Prediction (NWP) models is challenging due to the complex interaction between the influence of the low-level jet (LLJ), land-sea contrast, topography, and urban landscape \citep{Zhong2017,Yali2017,sun2002,xiarudi2006A,xia2009diagnosis,zhang20090112}. The land surface in the South China region is characterized by complex terrain and heterogeneity, which play a crucial role in promoting more active convections. Previous studies \citep{Giorgi2016,Saroj2018,Schumacher2020,Onishi2023} have shown that higher spatial resolution improves the performance of convective rainfall forecasts by resolving topographic characteristics more accurately. Recognizing the importance of resolution in forecasting severe convective weather, both the Chinese government and the community have shown increased support for developing high-resolution operational forecast models for warm-sector rainstorms and sudden local rainstorms. In early 2017, the China Meteorological Administration (CMA) initiated the development of a robust framework to evaluate the forecast accuracy of all available models, including high-resolution regional models, and to develop key technologies for high-impact weather forecasting.

With the increased availability computational resources, there has been a growing trend of using regional NWP models with finer grid spacings, typically ranging from 1 to 10 km. However, when the model grid spacing becomes comparable to the size of convection, which is known as the gray zone \citep{Wyngaard2004,Hong2012}, cumulus convection that was previously unresolved becomes partially resolved. In theory, unless Direct Numerical Simulation (DNS) is used to accurately capture the smallest turbulent scales with a resolution of millimeters to centimeters \citep{Julia2019}, parameterization of turbulence or convection remains necessary for weather modeling. Nevertheless, there is controversy about whether convection parameterization (CP) should be used in the gray zone. Some previous studies \citep{Chan2013,Johnson2013} reducing the horizontal grid spacing to below 4 km while using CP scheme does not yield any improvement, and in some cases even worsens precipitation forecast performance. In contrast, \citet{Schwartz2014} showed that forecasts with a horizontal grid spacing of 1 km provide a more accurate spatial representation of accumulated rainfall over 48 hours compared to forecasts with 4 km grid spacing. These conflicting findings typically arise when the CP scheme is applied at scales outside of its designed scales or when it is abruptly disabled at a resolution of approximately 3-5 km. Therefore, it remains unclear if utilizing any CP schemes in the gray zone is effective for predicting localized warm-sector heavy rainfall.

To enhance simulations in the gray zone, researchers hsave developed scale-aware CP schemes. These schemes dynamically parameterize convective processes based on the horizontal grid spacing and ensure a smooth transition across spatial scales. A study conducted by \citep{Julia2019} demonstrated that two scale-aware CP schemes, namely Grell-Freitas \citep{Grell2014} and multi-scale Kain-Fritsch (MSKF) \citep{Zheng2016}, outperform conventional CP schemes in terms of precipitation timing and intensity over the Southern Great Plains of the United States. Despite the growing use of these scale-aware schemes due to their superior performance, it is important to note that they also depend on various empirical parameters \citep{Pradas2022}. Hence, the development of CP schemes specific to the gray zone in NWP models still presents significant challenges.

In recent years, there has been a growing number of studies investigating the use of machine learning (ML) models as an alternative to conventional physics-based CP schemes. ML-based parameterization schemes have the potential to be effective at various horizontal resolutions because they are trained using data generated from models with varying grid resolutions \citep{yuval2020}. Unlike conventional CP schemes, which rely on assumptions like convective quasi-equilibrium \citep{Arakawa2004}, ML-based parameterization schemes do not depend on such assumptions. Random forests (RFs) and fully-connected (FC) neural networks (NNs) have been the two most frequently employed ML models for convective parameterization (CP) schemes in prior research. RFs automatically impose physical constraints, including energy conservation and non-negative surface precipitation, which are crucial for stable simulations. \citet{Gorman2018} demonstrated that RFs can maintain stability and accurately reproduce essential climate statistics by training them to mimic the moist convection of an aquaplanet general circulation model (GCM). More recently, \citet{yuval2020} employed the coarse-grained output of a high-resolution three-dimensional (3D) GCM model, simulated on an idealized equatorial beta plane, to train the RF parameterization. They showed that the RF parameterization is capable of reproducing the climate of the high-resolution simulation when used at coarse resolution. However, FC NNs offer several advantages over RFs, such as the potential for higher accuracy and reduced memory requirements. In their pioneering work, \citet{Krasnopolsky2013} formulated a stochastic CP scheme using an ensemble of 3-layer NNs, which were trained using data generated by a cloud-resolving model (CRM) over a small area in the tropical Pacific Ocean during a four-month period of the TOGA‐COARE \footnote{TOGA‐COARE is an acronym for Tropical Ocean Global Atmospheres/Coupled Ocean Atmosphere Response Experiment. It is an international research program that investigates the interaction or coupling of the ocean and atmosphere in the western Pacific warm pool region from November 1992 to February 1993, encompassing 120 days of field experiments involving the deployment of oceanographic ships, moorings, drifters, and Doppler radars (ship, land, air).}. The results revealed that the NN-based parameterization yielded reasonable and promising decadal climate simulations across a broader tropical Pacific region when incorporated into the National Center of Atmospheric Research (NCAR) Community Atmospheric Model (CAM). \citet{Gentine2018} utilized data from idealized simulations performed using the SuperParameterized Community Atmosphere Model (SPCAM) over an aquaplanet to train a deep NN (DNN). The DNN predicts temperature and moisture tendencies influenced by convection and clouds, along with the cloud liquid and ice water contents. Additionally, \citet{Rasp2018} successfully incorporated an NN‐based parameterization into a global GCM on an aquaplanet. They conducted stable prognostic simulations for multiple years, accurately reproducing the climatology of SPCAM and capturing crucial aspects of variability, including extreme precipitation and realistic tropical waves. However, \citet{Rasp2020} also found that minor modifications to the configuration rapidly led to unpredictable blow-ups in simulation. Therefore, addressing the instability of NN parameterization in GCMs is necessary. \citet{yuval2021} developed a FC NN to that predicts the subgrid fluxes instead of tendencies, incorporating the physical constraints from coarse-grained high-resolution atmospheric simulation in an idealized domain. \citet{Brenowitz2018,Brenowitz2019} proposed a novel loss function that minimizes the accumulated prediction error over multiple time steps instead of a single one. They further ensured long-term stability and accuracy by excluding upper atmospheric humidity and temperature from the inputs. However, the approach of removing particular variables from the inputs is relatively rudimentary, demanding additional research to enhance the stability of NN-based parameterizations when integrated into the model.

In addition, previous studies have mainly used FC NNs to emulate convection. However, there exist other advanced NN structures that have the potential to achieve higher accuracy. In a recent study, \citet{Han2020} made an initial attempt to employ a deep residual convolutional NN (ResNet) \citep{he2016deep} for emulating convection and cloud parameterization in the SPCAM model using a realistic configuration. They compared the performance of ResNet with other NN architectures, such as a FC DNN, a DNN with skip connections, and a convolutional NN (CNN) without skip connections. The results of their comparison demonstrated that both ResNet and CNN without skip connections outperformed the FC NN and the DNN with skip connections in terms of accuracy, with the performance of ResNet and CNN without skip connections exhibiting similar performance. This finding highlights the crucial role of convolutions in achieving higher accuracy. Another study conducted by \citet{YaoZhong2023} compared various ML model structures for emulating atmospheric radiative transfer processes, including FC NNs, CNNs, bidirectional recurrent-based NNs (RNNs), transformer-based NNs \citep{vaswani2017attention}, and Fourier Neural Operators (FNO \citep{li2020fourier}). Their results revealed that models with ability to preceive global context of the entire atmospheric column outperformed FC NNs and CNNs. Particularly, the bidirectional long short-term memory (Bi-LSTM) achieved the highest accuracy. Similar to radiative transfer modeling, \citet{Han2020} also emphasized the importance of ML having a global perspective of the entire atmospheric column for ML models in convection modeling. Modifying the depths of CNNs from 4 to 22 layers led to an increase in model accuracy. This improvement is primarily attributed to the expansion of the receptive field in deeper CNN layers. Thus, ML models that integrate both global and local perception capabilities are better suited for developing ML-based CP schemes.

Furthermore, all previous studies have predominantly focused on using CP schemes in GCM models for climate forecasting. Moreover, the choice of CP schemes significantly influences the uncertainty in precipitation forecasts within weather forecasting models. The complexity of the CP schemes also surpasses those applied in climate models \citep{Arakawa2004}. This study employs ML models to simulate convective processes for weather forecasting. We generate our dataset by running the Weather and Research Forecasting (WRF) \citep{Skamarock2021} model with the scale-aware MSKF scheme employed as the CP scheme. Developed as a means to mitigate the precipitation overestimation associated with the original KF scheme, the MSKF scheme serves as an improved version of the Kain-Fritsch (KF) scheme \citep{Kain1990,Kain1993,Kain2004}. The KF scheme often initiates convection prematurely, which leads to an overprediction of precipitation during summer. To address these issues, the MSKF incorporates a scale-dependent capability, such as modifying the formulation of the convective adjustment timescale. This vital parameter defines the intensity and duration of convection, and the MSKF scheme made it dynamic and grid resolution dependent \citep{Zhang2021}. The WRF model covers the South China region. Furthermore, we utilize a Bi-LSTM model to emulate the convective processes and couple it with the WRF model using the WRF-ML coupler developed by \citet{Zhong2023}. The performance of the ML-based CP scheme is evaluated in both offline and online settings.

The paper is structured as follows. Section 2 provides a description of the WRF model for data generation, as well as the input and output data of the ML model. In Section 3, the original and the ML-based MSKF scheme is introduced. The results for both offline and online testing of the ML-based MSKF scheme are presented in Section 4. Finally, Section 5 presents the summary and conclusion.

\section{Data}
\subsection{Data generation}
\label{data_generation}

The dataset was generated by running the WRF model version 4.3 \citep{skamarock2019description,Skamarock2021}. The following subsections provide a comprehensive explanation of the WRF model configurations, as well as the input and output variables employed in the development of the ML-based CP scheme.

\begin{figure}
\centering
\noindent\includegraphics[scale=0.08]{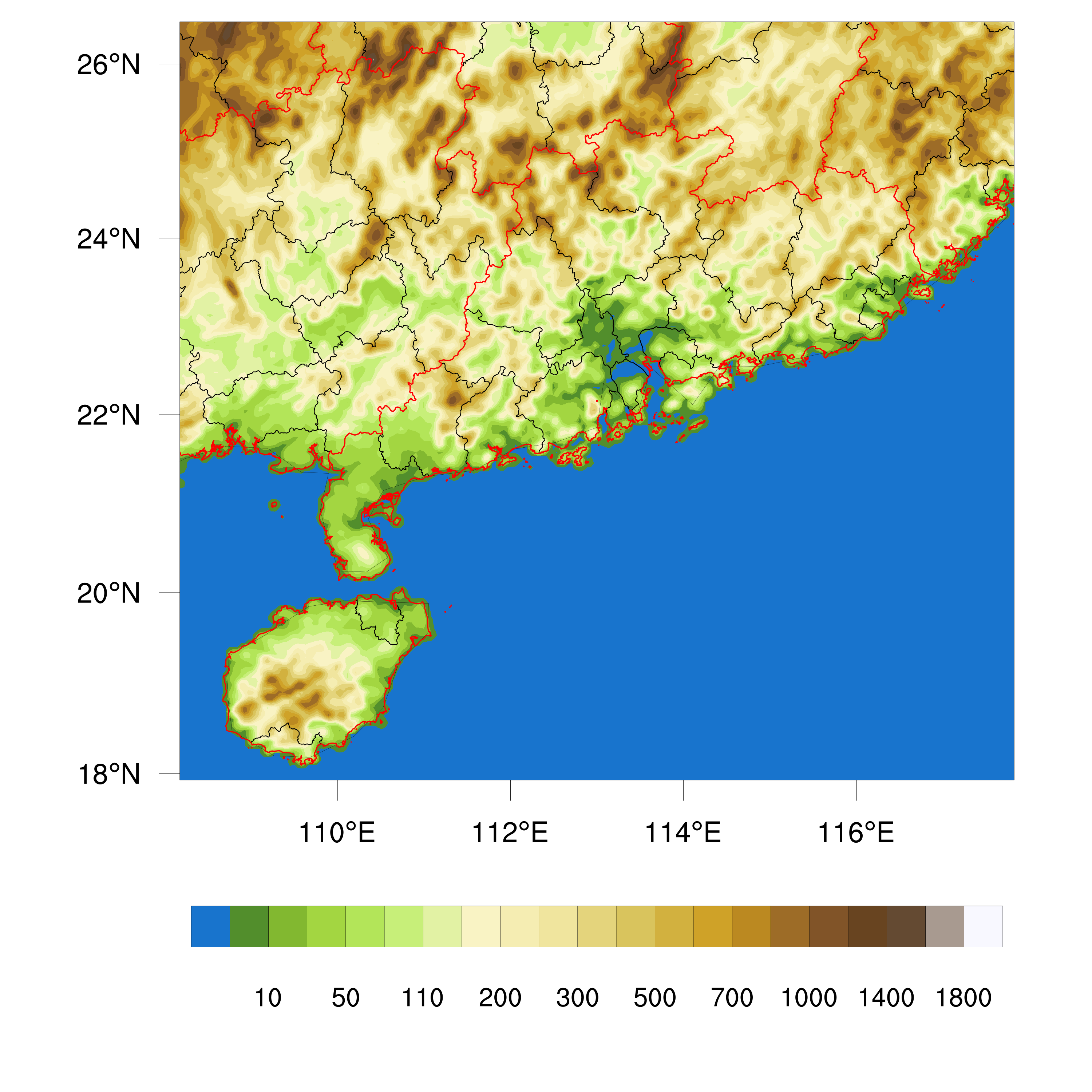}
\caption{Digital evaluation data of the single WRF domain with horizontal resolution at 5. Red lines are the province borderlines, and black lines are the city borderlines.}
\label{WRF_domain}
\end{figure}

The WRF model is compiled using the GNU Fortran (gfortran version 7.5.0) compiler with the "dmpar" option. The WRF model is run using the domain configuration illustrated in Figure \ref{WRF_domain}. The WRF model is configured with a single domain  consisting of 44000 grid points, with a horizontal grid spacing of 5 km and dimensions of 220 $\times$ 200 grid points in the west-east and north-south directions. The model consists of 45 vertical levels (i.e., 44 vertical layers), with a model top at 50 hPa. Additionally, the WRF model is configured with physics schemes, including WSM 6-class graupel scheme \citep{hong2006wrf} for microphysics, Bougeault-Lacarr{\`e}re (BouLac) scheme \citep{Bougeault1989} for planetary boundary layer (PBL) mixing, the Monin-Obukhov (Janjic) surface layer scheme \citep{janjic1996surface}, the Unified Noah model \citep{livneh2011development} for land surface, RRTMG for both shortwave and longwave radiation \citep{Iacono2008}, and MSKF \citep{Zheng2016} for cumulus. The time step used for all WRF simulations is set to 15 seconds.

The initial and boundary conditions were derived from the ERA5 reanalysis dataset, which was provided by the European Centre for Medium-range Weather Forecast (ECMWF) \citep{hersbach2020era5}. The ERA5 reanalysis dataset used in this study has a horizontal resolution of $0.25^\circ$ and consists of 29 pressure levels below 50 hPa. To create a dataset for developing the ML model, the WRF simulations were initialized at 12 UTC and conducted 9 times every 2 days, specifically from May 20th, 2022 to June 5th, 2022. Throughout the simulations, the MSKF scheme was called every 5 model minutes, generating outputs at each call. The simulations ran for 36 hours each time, with the first 24 hours used for training and the last 12 hours for validation. Therefore, the total number of training samples is 114,444,000 \footnote{114,444,000 = 44000 $\times$ 9 $\times$ (24 $\times$ 60 $\slash$ 5 + 1)} while the offline validation set contains 57,024,000 \footnote{57,024,000 = 44000 $\times$ 9 $\times$ 12 $\times$ 60 $\slash$ 5} samples.

Furthermore, considering that the offline performance might not necessarily reflect its performance in online setting, we conducted experiments by coupling the ML-based MSKF scheme with WRF model and comparing the results to the original WRF simulations to evaluate the online performance of the ML-based MSKF scheme. The simulations were conducted 4 times every 2 days, each lasting 36 hours. The initialization days spanned from June 12th, 2022 to June 18th, 2022. 

\FloatBarrier

\subsection{Input and output data}
\label{input_output}

Table \ref{table_variable} presents a comprehensive list of the input and output variables used in this study. There are 13 variables exclusively used as input, while 9 variables serve as both input and output. The output time-step precipitation due to convection ("raincv") is calculated through multiplying precipitation rate by the model time step. Out of all the variables, 5 are two-dimensional (2D) surface variables, while the remaining ones are 3D variables characterized by 44-layer vertical profiles. Additionally, the ML model used in this study incorporates 4 derived variables as input. These variables consist of a 2D Boolean variable indicating convection triggering based on the values of "nca", pressure difference between each adjacent vertical levels, saturated water vapor mixing ratio, and relative humidity. Furthermore, the output "w0avg", which depends on vertical wind component (w) and input "w0avg", is also included as an input to model. In total, the ML model utilizes 27 input variables.

The variable "nca" represents the cloud relaxation time and must be an integer multiple of the model time step. For all WRF model simulations conducted in this study, a fixed time step of 15 seconds is used. Thus, "nca" is expected to be evenly divisible by 15. To eliminate dependence on the specific model time step, "nca" is divided by the model time step before normalization is applied during model training. Moreover, within the MSKF scheme, "nca" plays a crucial role in determining the triggering of convection. Convection is triggered when "nca" is greater than or equal to half of the model time step.  

To ensure consistency with the dimensions of the 3D variables, the surface variables are padded by duplicating the values of the surface layer for all layers before feeding them into the model. Prior to utilizing the variables in the Bi-LSTM model for training and validation, normalization is applied to ensure uniformity in the magnitudes of all the variables. Each variable is divided by the maximum absolute value in the atmospheric column (for 3D variables) or at the surface (for surface variables). 

\begin{table}
\centering
\caption{Definition of all the input and output variables, and whether they are surface or 3D variables and their corresponding units. There are 44 model layers.}
\label{table_variable}
\small 
\begin{tabularx}{\textwidth}{ccXcc}
\hline
Type  & Variable name & Definition & type   & Unit     \\ 
\hline    
Input  & u & meridional wind component & 3D & m/s \\
       & v & zonal wind component & 3D & m/s \\
       & w & vertical wind component & 3D & m/s \\
       & t  & temperature & 3D & K  \\       
       & qv  & Water vapor mixing ratio  & 3D & kg/kg  \\
       & p & pressure & 3D & Pa \\
       & th & potential temperature & 3D & K \\       
       & dz8w & layer thickness & 3D & m \\           
       & rho & air density & 3D & kg/m$^{3}$ \\
       & pi & Exner function, which is dimensionless pressure and can be defined as: \((\frac{p}{p_0})^{R_d/c_p}\) &  \\       
       & hfx & upward heat flux at surface & surface & W/m$^{2}$ \\   
       & ust & u${*}$ in similarity theory & surface & W/m$^{2}$ \\   
       & pblh & planetary boundary layer height & surface & m \\        
\hline
Input and Output & rthcuten  & potential temperature tendency due to cumulus parameterization & 3D & K/s   \\
                 & rqvcuten  & water vapor mixing ratio tendency due to cumulus parameterization & 3D & kg/kg/s  \\
                 & rqccuten  & cloud water mixing ratio tendency due to cumulus parameterization & 3D & kg/kg/s  \\            
                 & rqrcuten  & rain water mixing ratio tendency due to cumulus parameterization & 3D & kg/kg/s  \\        
                 & rqicuten  & cloud ice mixing ratio tendency due to cumulus parameterization & 3D & kg/kg/s  \\        
                 & rqscuten  & snow mixing ratio tendency due to cumulus parameterization & 3D & kg/kg/s  \\
                 & w0avg  & average vertical velocity & 3D & m/s  \\                
                 & nca & counter of the cloud relaxation time & 3D & s \\
                 & pratec  & precipitation rate due to cumulus parameterization & surface &  mm/s \\
\hline
Output & raincv  & precipitation due to cumulus paramterization & surface &  mm \\                 
\hline
\end{tabularx}
\end{table}



\section{Method}
\label{method}

This section describes the flow chart of original MSKF scheme for determining convection trigger, ML model structures and training, and the evaluation methods.

\subsection{Description of original MSKF module}

\begin{figure}
\centering
\noindent\includegraphics[scale=0.7]{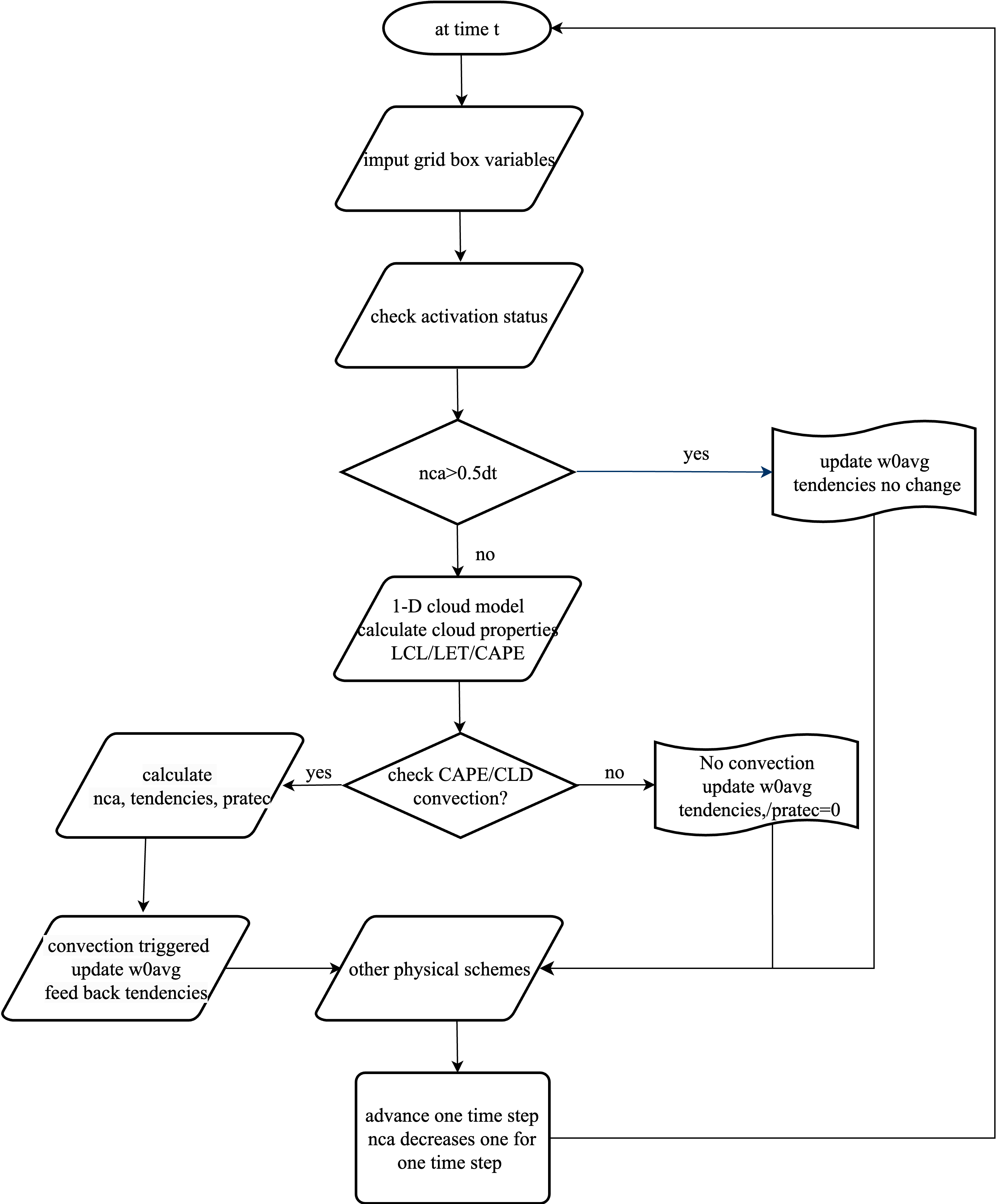}
\caption{A flow chart outlining convection trigger process in the original MSKF scheme.}
\label{MSKF_flow_chart}
\end{figure}

The MSKF scheme is a scale-aware adaptation of the KF CP scheme, which was originally developed by \citet{Kain1990,Kain1993} and later modified by \citet{Kain2004}. The flow chart illustrating the convection trigger process in the original MSKF scheme is illustrated in Figure \ref{MSKF_flow_chart}. At the beginning of each call, the variable "nca" is examined to determine if exceeds or equals a threshold equal to half of the model time step (dt). If "nca" is greater than or equal to half of dt, the convective tendencies and precipitation rate do not require updating since convection is still active. If "nca" is smaller than the specified threshold, a 1-D cloud model within the MSKF scheme is used to calculate various variables related to cloud properties in order to determine whether convective should be triggered. These variables include the lifting condensation level (LCL), convective available potential energy (CAPE), cloud top and base heights, and entrainment rates. If convection is activated for a particular grid point, the MSKF scheme computes the convective tendencies and precipitation rate. On the other hand, if convection is not triggered, the convective tendencies and precipitation rate remain at 0. However, the variable "w0avg" is always updated regardless whether convection is triggered or not. As long as the convection remains active, "nca" is decremented by one model time step at each WRF model's time step.

\subsection{Description of ML-based MSKF scheme}

In the original MSKF scheme, the atmospheric column is processed sequentially, one at a time, until all horizontal grid points within the domain have been processed. In contrast, the ML-based MSKF scheme performs calculations on a batch (B on Figure \ref{model}) of data, consisting of 44 features and 27 vertical layers. As a result, the input data has dimensions of B $\times$ 27 $\times$ 44. Before being fed into the ML model, the input data data undergoes pre-processing through a module incorporating a 1-dimensional (1D) convolutional layer. This module increases the feature dimension from 27 to 64. The subsequent sections provide a comprehensive description of the structures of the ML model.

\subsubsection{ML model structure}

\begin{figure}
\centering
\noindent\includegraphics[scale=0.7]{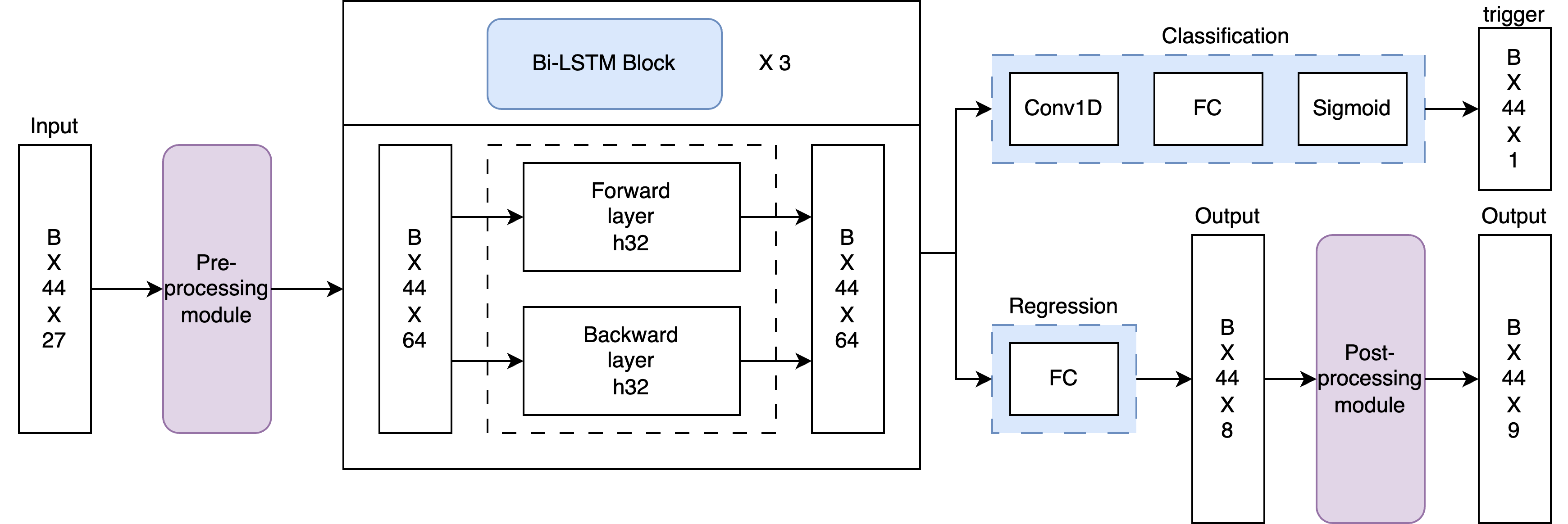}
\caption{The architecture of the multi-output Bi-LSTM model for combined classification and regression predictions.}
\label{model}
\end{figure}

Predicting whether convection is triggered as well as modeling convective tendencies and precipitaion rate are two main tasks in conventional CP schemes. Therefore, the development of a ML-based CP scheme requires both a binary classification model for predicting convection trigger and a regression model for modeling convective tendencies. Accordingly, we propose a multi-output Bi-LSTM model that can simultaneously perform regression and classification predictions (Figure \ref{model}). Our proposed model consists of a shared Bi-LSTM layer for learning features, a classification subnetwork, and a regression subnetwork. The shared Bi-LSTM layer includes three repeated Bi-LSTM blocks, with each block containing a forward and a backward layer that have a feature dimension of 32. The classification subnetwork is composed of a $1 \times 1$ 1D convolutional layer, a FC layer, and a Sigmoid activation layer. The output of the Sigmoid layer represents the probability distribution of the convection trigger. The binary cross-entropy loss function is employed as the cost function for this classification task. Meanwhile, the regression subnetwork incorporates a FC layer to output precipitation rate, "nca", and convective tendencies. Finally, the output of both subnetworks passes through a post-processing module to ensure physical consistency. The post-processing module is introduced in more detail in the following subsection.

\subsubsection{Post-processing module}
The post-processing module is designed to ensure physical consistency of all variables. The following rules are applied: 1) For grid points where the input "nca" is greater than or equal to half of dt, all other variables remain unchanged as they are still within the convection lifetime. 2) The output "nca" must be an integer. 3) For grid points where convection is predicted to be inactive, all other output variables are set to zero. In addition, the time-step convective precipitation (raincv) is calculated as described in the previous section \ref{input_output}.

\subsubsection{Model training}

As our model incorporates both classification and regression tasks, we optimize its performance by minimizing a multi-task loss function \citep{ren2016faster}. The loss function is defined as the sum of the binary cross entropy loss for the convection trigger and a weighted combination of the ${L1}$ loss for all output variables from the regression subnetwork. The specific formulation of the loss function is as follows:
\begin{equation} 
\label{loss}
L =  \frac{1}{N_{cls}} \sum_{i,j} L_{cls}(p_{i,j}, p_{i,j}^{gt}) + \sum_{c} \lambda_{c} \frac{1}{N_{reg}} \sum_{i,j} p_{i,j}^{gt} L1_{c}
\end{equation}
Here, $i$ and $j$ denote the grid points in the domain. ${p_{i,j}}$ represents the probability of convection being triggered. The ground-truth label $p_{i,j}^{gt}$ takes a value of 1 if convection is triggered and 0 otherwise. The classification loss $L_{cls}$ is calculated using the binary cross entropy loss. For the regression loss of different variables ${c}$, ${\lambda_{c}}$ functions as a weight that balances the output variables by considering their respective magnitudes. The term ${p_{i,j}^{gt} L1_{c}}$ indicates that the ${L1}$ regression loss is activated only for triggered grid points (${p_{i,j}^{gt} = 1}$) and is disabled otherwise (${p_{i,j}^{gt} = 0}$). Both loss terms are normalized by ${N_{cls}}$ and ${N_{reg}}$, which correspond to the total number of grid points and the number of triggered grid points, respectively.

Adam optimizer \citep{kingma2014} is used with an initial learning rate of 0.003 update the parameters of the model. Furthermore, the plateau scheduler is implemented to decrease the learning rate by a factor of 0.5 when the loss fails to decrease for five epochs. The model is trained for 150 epochs using a batch size of 44000.

\subsection{Evaluation methods}
The ML-based MSKF scheme is evaluated in both offline and online settings. The offline performance of the ML-based MSKF scheme is evaluated by comparing it against the outputs of the original MSKF scheme using validation dataset, including rthcuten, rqvcuten, rqccuten, rqrcuten, nca, and pratec. The overall model performance metrics include root mean squared error (RMSE) and correlation coefficient. The mean absolute error (MAE) and mean bias error (MBE) per vertical layer were were calculated using the equation below:

\begin{equation} \label{MAE}
MAE_l = \frac{1}{N}\sum_{i=1}^{N} \arrowvert Y_{ML}(i,l) - Y(i,l) \arrowvert
\end{equation}

\begin{equation} \label{MBE}
MBE_l = \frac{1}{N}\sum_{i=1}^{N}  Y_{ML}(i,l) - Y(i,l)
\end{equation}
where $Y(i,l)$ and $Y_{ML} (i,l)$ represent the outputs from the original MSKF scheme and ML-based MSKF scheme, respectively. Here, $i$ denotes the horizontal grid point of a vertical profile, $N$ is the number of the horizontal grid points in the domain, $l$ represents the vertical layer index.

\section{Results}
\label{results}

\subsection{Offline validation of the ML-based MSKF scheme}

\begin{figure}
\noindent\includegraphics[scale=0.45]{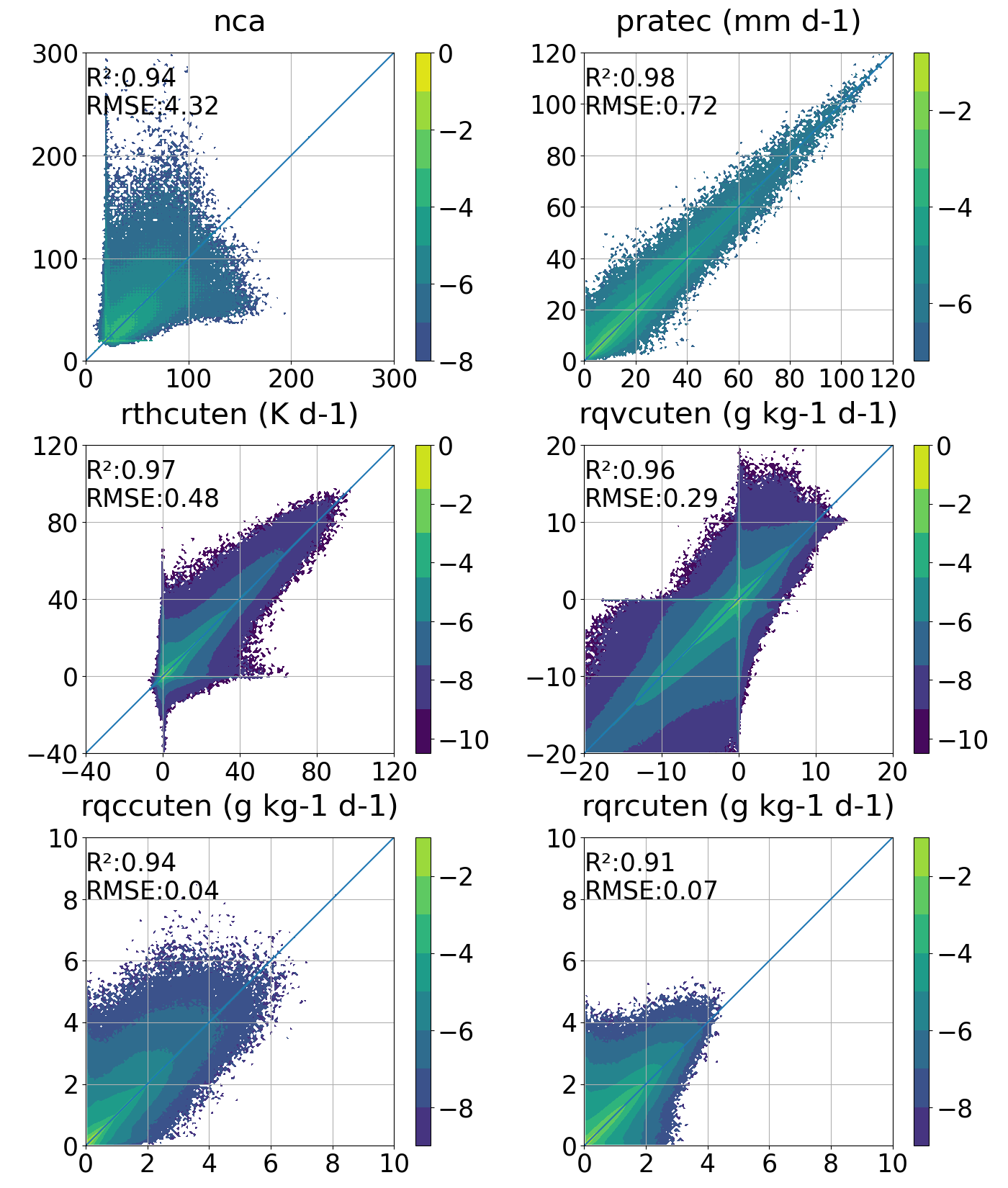}
\caption{Comparison of the predicted (y axis) and true (x axis) nca, pratec, rthcuten (first column), rqvcuten (second column), rqccuten (third column), and rqrcuten for using validation data in the offline setting.}
\label{correlation}
\end{figure}

The offline validation was conducted using data that was not used during the training process. Figure \ref{correlation} compares the cloud relaxation time (nca), precipitation rate (pratec), and convective tendencies predicted by both the original MSKF scheme and the ML-based MSKF scheme, respectively.To facilitate the comparison, the precipitation rate and temperature tendencies were multiplied by a factor of 86,400 (24 $\times$ 3600) to convert from mm·s$^{-1}$ and K·s$^{-1}$ to mm·d$^{-1}$ and K·d$^{-1}$, respectively. Similarly, the water vapor mixing ratio (rqvcuten), cloud water mixing ratio (rqccuten), and rain water mixing ratio (rqrcuten) resulting from convection were multiplied by 86,400,000 (24 $\times$ 3600 $\times$ 1000) to convert from kg·kg$^{-1}$·s$^{-1}$ to g·kg$^{-1}$·d$^{-1}$. Among the output variables listed in Table \ref{table_variable}, w0avg is excluded as it is calculated using an equation with the ground truth as input in this offline validation. Hence, evaluating w0avg in the offline evaluation is unnecessary.

Among all the variables illustrated in Figure \ref{correlation}, the variable "nca" exhibits a significantly higher RMSE of 4.32, and its data points appear scattered across a wide range of values. This suggests that accurately predicting convection poses a considerable challenge. Prior to plotting and performing statistical calculations, "nca" is divided by the model time step of 15 to eliminate the time step dependence. The precipitation rate demonstrates the highest correlation coefficient and the smallest spread, as most data points cluster closely around the 1:1 line. In the case of temperature and the four moisture tendencies, the data also displays some dispersion, but a majority of the data points remain close to the 1:1 line. Overall, the ML-based MSKF scheme shows a strong correlation with the original MSKF scheme for all the variables, with correlation coefficients consistently higher than 0.91. This indicates that the the ML-based MSKF scheme has the potential to replace the original scheme. 

\begin{figure}
\noindent\includegraphics[scale=0.25]{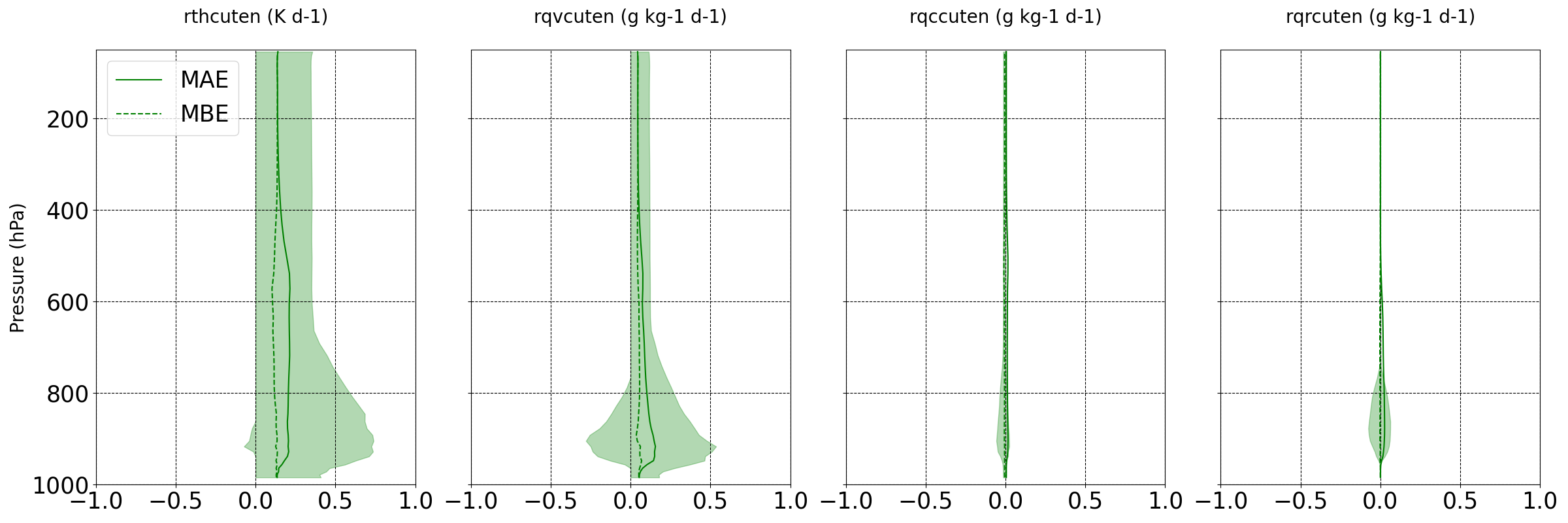}
\caption{Vertical profiles of the statistics in rthcuten (first column), rqvcuten (second column), rqccuten (third column), and rqrcuten (fourth column) using validation data in the offline setting data using ML-based emulators. The solid and dotted lines show the MAE and MBE profile, respectively, and the shaded area indicates the 5th and 95th percentile of differences (prediction—target) at each layer.}
\label{height_distribution}
\end{figure}

To gain a comprehensive understanding of the vertical distribution of errors, Figure \ref{height_distribution} illustrates the vertical profiles of statistics in convective tendencies. The solid and dotted lines in the figure represent the MAE and MBE of tendencies at each vertical layer, respectively. Additionally, the shaded area corresponds to the 5th and 95th percentiles of differences between tendencies predicted by the ML-based MSKF predicted scheme and the original MSKF scheme, respectively. The vertical error distribution in all tendencies is quite similar with higher variance observed among the pressure layers between 800 and 1,000 hPa. These pressure layers corresponds to the atmospheric layer where convection occurs most frequently. Due to the significantly lower cloud and rain content compared to water vapor in the atmosphere, the error magnitudes for rqccuten and rqrcuten are considerably smaller than those for rqvcuten.

\FloatBarrier

\FloatBarrier

\subsection{Prognostic validation}
This subsection presents the performance of the ML-based MSKF scheme in the online setting.

The ML-based MSKF scheme was integrated as a replacement for the original MSKF scheme in the WRF model to model convective effects. The WRF-ML coupler \citep{Zhong2023} was used to incorporate the ML-based MSKF scheme into the WRF model. Both the modified WRF model, incorporating the ML-based scheme, and the original WRF model were initialized on June 12, 14, 16, and 18, 2022, and run for 36 hours. It is worth mentioning that this runtime was completely independent of the training dataset.

\begin{figure}
\noindent\includegraphics[scale=0.17]{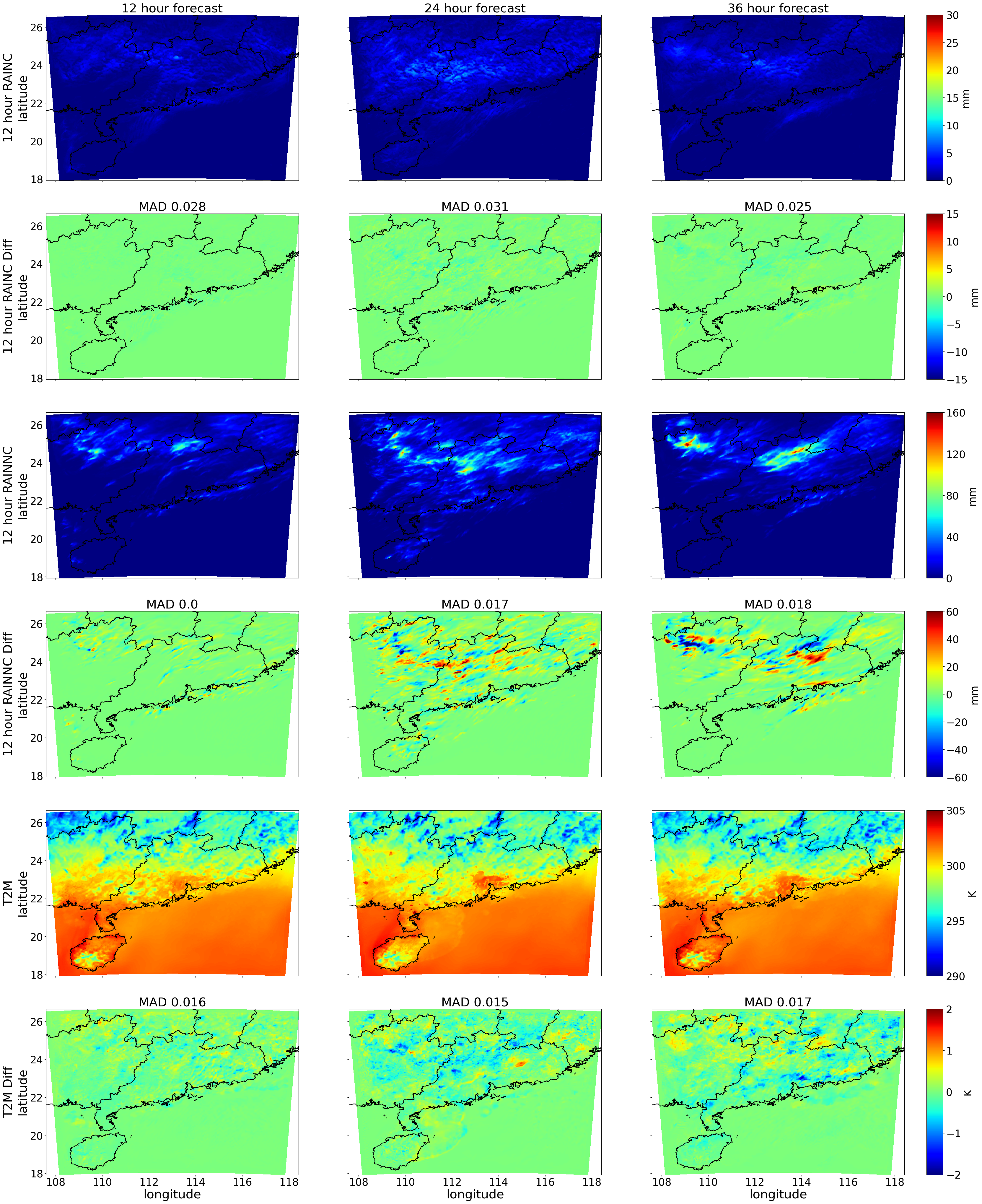}
\caption{Spatial map of the average WRF simulations using the original MSKF scheme (in the first, third, and fifth rows) along with the average MAD between WRF simulations coupled with the ML-based MSKF scheme and WRF simulation with the original MSKF scheme (in the second, fourth and sixth rows). The simulations are shown for the 12-hour accumulated convective precipitation (${RAINC}$) in the first and second rows, the 12-hour accumulated non-convective precipitation (${RAINNC}$) in the third and fourth rows, and the 2-meter temperature (${T2M}$) at forecast lead times of 12 hours (first column), 24 hours (second column), and 36 hours (third column).}
\label{spatial_map}
\end{figure}

In Figure \ref{spatial_map}, the averaged spatial forecasts of the 4-day prediction from the original WRF model are presented. These results include accumulation of convective and non-convective precipitation over a 12-hour period, as well as the 2-meter temperature at 12, 24, and 36 hours. The figure also illustrate the mean absolute difference (MAD) between the WRF simulations coupled with the ML-based MSKF scheme and those using the original MSKF scheme. The red and blue patterns in the spatial forecasts represent the magnitudes of the forecast values. In the spatial difference, red and blue patterns reveal the positive and negative biases of the ML-based simulations, respectively. while green patterns suggest little to no difference compared to the original WRF simulations. Additionally, a domain-averaged MAD is calculated to evaluate the overall performance of the ML-based scheme in prognostic simulations. Generally, the difference is small suggesting that the WRF simulations coupled with ML-based MSKF scheme agree well with the origianl WRF simulations. Also, the difference does not increase as the simulation time progresses, as there is a comparable domain-averaged MAD at 36 forecast hours compared to that at 12 forecast hours. These findings indicate that the ML-based MSKF scheme achieves stable prognostic simulations. 

\FloatBarrier

\conclusions  
\label{conclusions}
In this paper, we proposed a multi-output Bi-LSTM model to developing a ML-based MSKF scheme for predicting convection trigger and reproducing the convective process in the gray zone. The model is trained using data generated from the WRF simulations at a spatial resolution of 5 km, covering the South China region. The output variables of the ML-based MSKF scheme are identical to those of the original MSKF scheme, which include cloud relaxation time ("nca"), precipitation rate ("pratec"), time-step convective precipitation ("raincv"), and convective tendencies. In the ML-based scheme, the physical consistency among all output variables is considered and encoded as a post-processing module to refine the output from the Bi-LSTM model. Offline validation demonstrates the excellent performance of the ML-based MSKF scheme. Furthermore, the ML-based MSKF scheme is coupled with the WRF model using WRF-ML coupler. The WRF simulations coupled with the ML-based MSKF scheme is compared against the WRF simulation with the original MSKF scheme. Results shows that the ML-based scheme can generate forecasts similar to the original ML scheme in online settings, showing the potential substitution of the MSKF scheme by ML models in gray-zone.





\codedataavailability{The source code for the WRF model version 4.3 used in this study is available at \url{https://doi.org/10.5281/zenodo.10039053} \citep{skamarock_WRF}. The source code and data used in this are available at \url{https://doi.org/10.5281/zenodo.10032404} \citep{code2023}.} 












\authorcontribution{X.Z. trained the deep learning models and calculate the statistics of model performance. X.Y. and X.Z. conducted the WRF simulations to provide dataset for training and evaluation, and offered valuable suggestions on the model training and paper revision. X.Z. and X.Y. wrote, reviewed and edited the original draft; X.Z., X.Y., and H.L. supervised and supported this research and gave important opinions. All of the authors have contributed to and agreed to the published version of the manuscript.} 

\competinginterests{The authors declare no conflict of interest.} 


\begin{acknowledgements}
This work was supported by Basic and Applied Basic Research Foundation of Guangdong Province, under Grant No. 2021A1515012582. We are thankful for Mesoscale and Microscale Meteorology Laboratory (MMM) at NCAR for developing and sharing the WRF source codes.

\end{acknowledgements}


\bibliographystyle{copernicus}
\bibliography{refs}

\end{document}